\documentclass{sig-alternate}
\usepackage{array}

\usepackage{multirow}
\usepackage{tabulary}
\usepackage{subcaption}

\graphicspath{{figs/}}

\usepackage{caption}
\usepackage[caption=false,font=footnotesize]{subfig}
\usepackage{fixltx2e}
\usepackage{url}

\newcommand{\wildcard}{{\boldsymbol ?}}
\newcommand{\gap}{{\boldsymbol \star}}
\newlength{\aligncharw}
\newcommand{\agap}{\makebox[\aligncharw][s]{\scriptsize{$\gap$}}}
\newcommand{\eg}{e.g.\,}
\newcommand{\ie}{i.e.\,}
\newcommand{\cf}{cf.\,}

\newcommand{\prototype}{prototype}
\newcommand{\mprototype}{message prototype}
\newcommand{\Prototype}{Prototype}

\newcommand{\refFigure}[1]{Figure \ref{#1}}

\begin{document}
%
\title{Opaque Service Virtualisation: A Practical Tool for Emulating Endpoint Systems}



%
\numberofauthors{6} 
%
\author{
%
%
\alignauthor
Steve Versteeg\\
       \affaddr{CA Research}\\
       \affaddr{CA Technologies}\\
       \affaddr{Melbourne, Australia}\\
       \email{steve.versteeg@ca.com}
\alignauthor
Miao Du\\
       \affaddr{Swinburne University of Technology}\\
       \affaddr{Hawthorn, Victoria, Australia}\\
       \email{miaodu@swin.edu.au}
\alignauthor Jean-Guy Schneider\\
       \affaddr{Swinburne University of Technology}\\
       \affaddr{Hawthorn, Victoria, Australia}\\
       \email{jschneider@swin.edu.au}
\and
\alignauthor John Grundy\\
       \affaddr{Deakin University}\\
       \affaddr{Burwood, Victoria, Australia}\\
       \email{j.grundy@deakin.edu.au}
\alignauthor Jun Han\\
       \affaddr{Swinburne University of Technology}\\
       \affaddr{Hawthorn, Victoria, Australia}\\
       \email{jhan@swin.edu.au}
\alignauthor Menka Goyal\\
       \affaddr{CA Technologies}\\
       \affaddr{Plano, Texas, USA}\\
       \email{menka.goyal@ca.com}
}



\maketitle


\begin{abstract}
\boldmath
Large enterprise software systems make many complex interactions with other services in their environment. Developing and testing for production-like conditions is therefore a very challenging task. Current approaches include emulation of  dependent services using either explicit modelling or record-and-replay approaches. Models require deep knowledge of the target services while record-and-replay is limited in accuracy. Both face developmental and scaling issues. We present a new technique that improves the accuracy of record-and-replay approaches, without requiring prior knowledge of the service protocols. The approach uses Multiple Sequence Alignment to derive {\mprototype}s from recorded system interactions and a scheme to match incoming request messages against {\prototype}s to generate response messages. We use a modified Needleman-Wunsch algorithm for distance calculation during message matching. 
Our approach has shown greater than 99\% accuracy for four evaluated enterprise system messaging protocols. The approach has been successfully integrated into the CA Service Virtualization commercial product to complement its existing techniques.
%
%
\end{abstract}



%

\section{Introduction}
\label{intro.sec}

Large enterprise software systems must be rigorously tested before deployment. However, modern software systems are becoming ever increasingly inter-connected.
In a typical deployment scenario, an enterprise system might interact with many other disparate systems, such as mainframes, directory servers, databases, and other types of software services. In order to check whether the enterprise system will function correctly in terms of its interactions with these services, it is necessary to test it in as realistic an environment as possible, prior to the updated system's actual deployment.  In the increasingly popular DevOps (Development-Operations) environments~\cite{bass:2015} where changes are continuously pushed out to users, this testing needs to be carried out both rigorously but also quickly and often repeatedly in a short period.

Getting access to the actual production environment for testing is not possible due to the risk of disruption.  Large organisations often have a test environment, which is a close replication of their production environment, but this is both very expensive to set up and use.  Furthermore the test environment is in high demand, so software developers will have only limited access to it, which may greatly slow releases.  Enabling developers to have continuous access to production-like conditions to test their application is an important part of DevOps but is as yet an unsolved problem. One popular approach, to test an application's dependence on other systems, is to install the other systems on virtual machines (such as VMware) \cite{Sugerman:01}.  However, virtual machines are time consuming to configure and maintain.  Furthermore, the configuration of the systems running on the virtual machine is likely to be different to the production environment. 

An alternative that is gaining traction is service emulation (also called
service virtualisation), where models of services are emulated to provide more
realistic scale and less complicated configuration\cite{hine:thesis}. However,
existing approaches to service emulation require detailed knowledge of the
target service's protocol and message structure. Emulation tools provide
\emph{data protocol handlers} for built-in support of commonly used
protocols, but it is not possible to support every protocol used by every
application. Where a target service uses a protocol not supported by the
emulation tool, then system experts need to explicitly model the target
services.  This is often infeasible if the required knowledge is unavailable
for the wide range of services in a real
deployment environment, especially legacy services, \cite{Ghosh99issuesin}, and is very time consuming and error-prone.

In this paper we describe our work in developing a practical, scalable
and fully automated approach to service emulation which uses no explicit
knowledge of the services, their message protocols and structures, and yet can
simulate -- to a high degree of accuracy and scale -- a realistic enterprise
system deployment environment.  The approach achieves very high assurance, real-time
performance, and is robust under varying message encodings, operation types
and payloads. Our focus is on generating service responses to enterprise
system requests, not models of the target services, unlike most other work in
this domain \cite{hine:thesis}. Our approach can
very often successfully emulate a response to a system-under-test request, to
support its testing under realistic deployment
conditions. The technique thus compliments rather than replaces other
conformance and integration testing techniques.

Our approach, called \emph{opaque service virtualisation}, is based on analysing message traces. The method extracts the descriptive {\prototype}s for each type of protocol operation. These prototypes form the basis of an opaque service model which can be used at runtime to send back the real-time responses to live requests coming from a system-under-test.



%

We describe our experience in taking our research proof-of-concept implementation and re-engineering it for incorporation into the CA Technologies' commercial product, CA Service Virtualization. Our technique has been successfully integrated into Service Virtualization and deployed to thousands of users.


\section{Related Work}
\label{sec:relatedwork}

The most common approach to providing a testing environment for enterprise systems using such a messaging protocol is to use virtual machines \cite{li:10}. Implementations of the services are deployed on virtual machines and communicated with by the system under test. Major challenges with this approach include configuration complexity \cite{vm4testing} 
 and the need to maintain instances of each and every service type in multiple configurations \cite{grundy:05}. Recently, cloud-based testing environments~\cite{testenvcloud} as well as containerisation~\cite{docker} have emerged to mitigate some of these issues.

Emulated testing environments for enterprise systems, relying on service
models, is another approach. When sent messages by the system
under test, the emulation responds with approximations of ``real'' service response
messages \cite{servicemodel}. Kaluta \cite{hine:thesis} is proposed to
provision emulated testing environments. Challenges with these approaches
include developing the models, lack of precision in the models, especially
for complex protocols, and ensuring robustness of the models under diverse
load conditions \cite{sun2012usefulness}.  To assist developing reusable
service models, approaches either reverse engineer message structures
\cite{cui:07a}, or discover processes \cite{processmining}. 
While these allow engineers to
develop more precise models, none of them can automate the creation
of executable interactive models of the communication between
a system under test and the dependency services.

Recording and replaying message traces is an alternative approach. This involves recording request messages sent by the system under test to real services and response messages from these services, and then using these message traces to `mimic' the real service response messages in the emulation environment \cite{cui:06}. Some approaches combine record-and-replay with reverse-engineered service models. CA Service Virtualization~\cite{Michelsen:12} is a commercial software tool, which can emulate the behaviour of services. The tool uses built-in knowledge of some protocol message structures to model services and mimic interactions automatically. However, current approaches have limited robustness, accuracy and performance.

In contrast, our goal is to produce an emulation environment for enterprise system testing
that uses message trace recordings collected a priori to produce a response on
behalf of a service when invoked by a system-under-test at runtime that is fast, accurate and has no knowledge of the underlying protocols.

%
%
%
%

\section{Approach}
\label{sec:approach}

\begin{figure}[h]
     \centering
     \includegraphics[width=8.6cm]{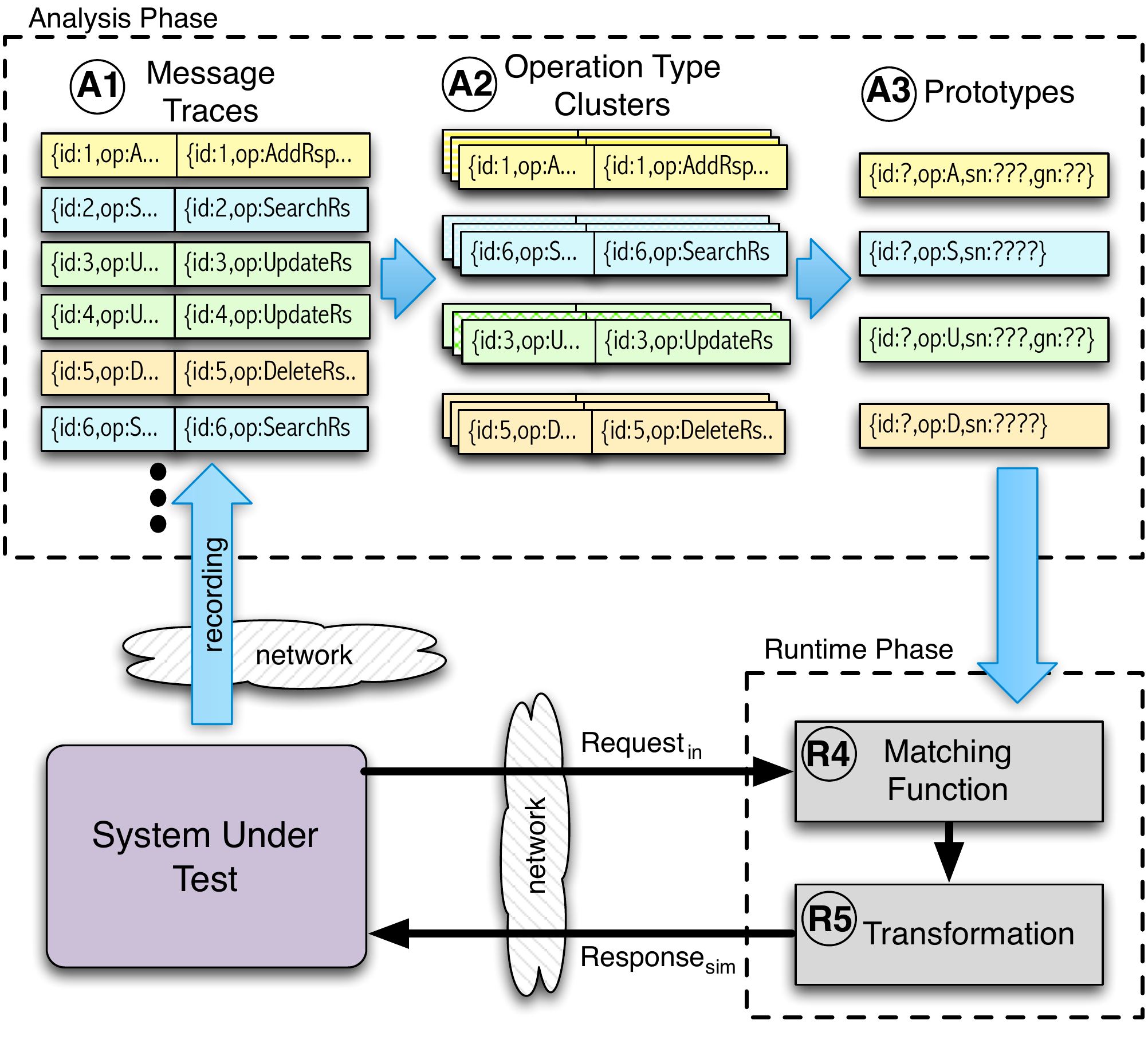}
\vspace{-.5cm}
     \caption{System overview}
     \label{fig:system_overview}
\end{figure}

The approach we take is to cluster the trace recordings into groups of similar
messages and then formulate a single representation -- a {\em \prototype}
capturing the common features -- for the request messages of each cluster.
This accelerates runtime performance by enabling incoming requests from the
system under test to be compared only to the prototypes rather than every
recorded request. Our approach (see \refFigure{fig:system_overview}) consists
of two stages: an analysis phase which is performed offline, and a runtime (or
playback) phase.

\smallskip

The analysis phase includes three steps:

\begin{description}
\vspace{-0.2cm}
\item[A1] We collect message traces of communications be- tween a
client and the real target service and store them in a \emph{transaction
library}.
\vspace{-0.2cm}
\item[A2] We cluster the transaction library, with the goal of
grouping transactions (request-response pairs) by operation type. We
do not consider the state of service under which the requests are
issued.
(This may give lower accuracy but is still useful in many cases as discussed in Section~\ref{sec:discussion})
\vspace{-0.2cm}
\item[A3] We derive a \emph{request \prototype} for each operation type cluster by performing a multiple sequence alignment and extracting common patterns.
\end{description}

\vspace{-0.2cm}
A further two steps are performed at runtime:
\begin{description}
\setcounter{enumi}{3}
\vspace{-0.2cm}
\item[R4] For an incoming live request message received from the system-under-test, we use \emph{a matching distance calculation} technique to select the nearest matching request {\prototype}.
\vspace{-0.2cm}
\item[R5] We perform dynamic substitutions on a specifically chosen response message from the identified operation type cluster to generate a modified response message sent back to the system-under-test.
\end{description}


\subsection{Needleman-Wunsch}
\label{ss:nw}

The Needleman-Wunsch algorithm \cite{needleman:1970} is a dynamic programming
algorithm for computing the edit distance between two sequences. We use it at different
steps during both the analysis and runtime phases. 
Needleman-Wunsch finds the globally optimal alignment for two sequences of
symbols in $O(m\! \cdot\! n)$ time, where $m$ and $n$ are the lengths of the
sequences. The algorithm uses a progressive scoring function $S$, which gives
an incremental score for each pair of symbols in the alignment. A different
score will be given depending on whether the symbols are identical, different,
or a gap has been inserted.

\subsection{Capturing Message Traces (Step A1)}
\label{ss:capture}

\begin{table*}[t]
\begin{center}
\begin{tabular}{|c||l|l|}
\hline
Index & Request & Response \\ \hline\hline
1 & \{id:1,op:S,sn:Du\} & \{id:1,op:SearchRsp,result:Ok,gn:Miao,sn:Du,mobile:5362634\} \\ \hline
13 & \{id:13,op:S,sn:Versteeg\} & \{id:13,op:SearchRsp,result:Ok,gn:Steve,sn:Versteeg,mobile:9374723\} \\ \hline
24 & \{id:24,op:A,sn:Schneider,mobile:123456\} & \{id:24,op:AddRsp,result:Ok\} \\ \hline
275& \{id:275,op:S,sn:Han\} & \{id:275,op:SearchRsp,result:Ok,gn:Jun,sn:Han,mobile:33333333\} \\ \hline
490 & \{id:490,op:S,sn:Grundy\} &
\{id:490,op:SearchRsp,result:Ok,gn:John,sn:Grundy,mobile:44444444\} \\ \hline
2273 & \{id:2273,op:S,sn:Schneider\} & \{id:2273,op:SearchRsp,result:Ok,sn:Schneider,mobile:123456\} \\ \hline
2487 & \{id:2487,op:A,sn:Will\} & \{id:2487,op:AddRsp,result:Ok\} \\ \hline
3106 & \{id:3106,op:A,sn:Hine,gn:Cam,Postcode:33589\} & \{id:3106,op:AddRsp,result:Ok\} \\
\hline
\end{tabular}
\end{center}
\vspace{-.5cm}
\caption{Directory Service Interaction Library Example}
\label{tab:tl}
\end{table*}

The first step is to record real message exchanges between a client
(such as a previous version of the system under test) and the service we aim to emulate,
and store them in a \emph{transaction library}.
A \emph{transaction} records a request sent by the client and the responses (zero or more)
returned by the service.
We capture transactions at the network level
(using a tool such as Wireshark) to record the bytes in TCP packet
payloads. We thereby make no assumptions
about the service message format.
Table~\ref{tab:tl} shows a small example transaction library
which we use in the following sections to illustrate how our method works.
It is from a fictional directory service protocol that has some similarities to the widely used
LDAP protocol~\cite{ldap}, but is simplified to make our running example easier to follow.
Our example protocol uses a JSON encoding. The transaction library contains two
kinds of operations: {\sf {\small add}} and {\sf {\small search}}. Add requests contain the field `op:A', whereas
search requests have `op:S'. The add and search response operations are specified with the fields
`op:AddRsp' and `op:SearchRsp', respectively.

\subsection{Clustering Transactions (Step A2)}

Having recorded our transaction library, the next step is to group the
transactions by operation type, but again without assuming any
knowledge of the message format.  To achieve this we use a distance function-based clustering technique.
In our previous work \cite{Du:2013SoftMine} we considered multiple cluster distance functions and found that the response similarity (as measured by edit distance, calculated using the Needleman-Wunsch algorithm \cite{needleman:1970}) was the most effective method for grouping transactions
of the same operation type.  We therefore use the same technique in this work, grouping
transactions by the similarity of their response messages.  The clustering algorithm used was VAT
\cite{VAT}, one among many alternatives, as it was effective in our previous work \cite{Du:2013SoftMine}.


Clustering the example transaction library
yields two clusters, as shown in tables \ref{tab:searchcluster} and
\ref{tab:addcluster}, corresponding to the {\sf {\small add}} and {\sf {\small
    search}} operations, respectively.

%

\begin{table*}[h]
\begin{center}
\begin{tabular}{|c||l|l|}
\hline
Index & Request & Response \\ \hline\hline
1 & \{id:1,op:S,sn:Du\} & \{id:1,op:SearchRsp,result:Ok,gn:Miao,sn:Du,mobile:5362634\} \\ \hline
13 & \{id:13,op:S,sn:Versteeg\} & \{id:13,op:SearchRsp,result:Ok,gn:Steve,sn:Versteeg,mobile:9374723\} \\ \hline
275& \{id:275,op:S,sn:Han\} & \{id:275,op:SearchRsp,result:Ok,gn:Jun,sn:Han,mobile:33333333\} \\ \hline
490 & \{id:490,op:S,sn:Grundy\} &
\{id:490,op:SearchRsp,result:Ok,gn:John,sn:Grundy,mobile:44444444\} \\ \hline
2273 & \{id:2273,op:S,sn:Schneider\} & \{id:2273,op:SearchRsp,result:Ok,sn:Schneider,mobile:123456\} \\ \hline
\end{tabular}
\end{center}
\vspace{-.5cm}
\caption{Cluster 1 (search operations)}
\label{tab:searchcluster}
\end{table*}

\begin{table*}[!h]
\begin{center}
\begin{tabular}{|c||l|l|}
\hline
Index & Request & Response \\ \hline\hline
24 & \{id:24,op:A,sn:Schneider,mobile:123456\} & \{id:24,op:AddRsp,result:Ok\} \\ \hline
2487 & \{id:2487,op:A,sn:Will\} & \{id:2487,op:AddRsp,result:Ok\} \\ \hline
3106 & \{id:3106,op:A,sn:Hine,gn:Cam,postalCode:33589\} & \{Id:3106,Msg:AddRsp,result:Ok\} \\
\hline
\end{tabular}
\end{center}
\vspace{-.5cm}
\caption{Cluster 2 (add operations)}
\label{tab:addcluster}
\end{table*}

\subsection{The Request \Prototype~(Step A3)}


The core contribution of our approach is a method to formulate representative
{\prototype}s. This step consists of three major subparts: (i) aligning the
common features of the request messages in a cluster, (ii) extracting the
common features while accommodating variations to derive the prototype, and
(iii) weighting sections of the prototype according to their relative
importance through an entropy analysis.

\subsubsection{Multiple Sequence Alignment}
\label{ss:msa}
In aligning the request messages of a cluster,
we adopt the multiple sequence alignment (MSA) technique~\cite{wang1994complexity} originated from 
bioinformatics. MSA was first used to align three or more biological sequences to reveal their structural commonalities~\cite{book:biologicalsequenceanalysis}\footnote{For this reason this technique has
also been widely used to reverse-engineer protocol message
structures~\cite{Comparetti:2009}.}.
Specifically, we adopt \emph{ClustalW}~\cite{clustalw}, a widely used heuristic technique
for MSA. It is memory efficient and is shown to produce high accuracy
alignments in polynomial computation time for empirical datasets (in contrast to the original NP-complete MSA technique~\cite{wang1994complexity}).  

Figure~\ref{chap7fig:alignment} shows the multiple sequence alignment
results of applying the $ClustalW$ algorithm to the example clusters from tables
\ref{tab:searchcluster} and \ref{tab:addcluster}.  The MSA results are
known as \emph{profiles}.
Gaps which were inserted during the alignment
process are denoted by the `$\gap$' symbol.

\newcommand{\alignedSrqs}{
	\begin{minipage}[b]{.45\linewidth}
	\texttt{\footnotesize{
	\{id:\agap{\agap}1{\agap},op:S,sn:{\agap}{\agap}{\agap}{\agap}{\agap}{\agap}{\agap}Du\} \linebreak
	\{id:{\agap}{\agap}13,op:S,sn:{\agap}Versteeg\} \newline
	\{id:2273,op:S,sn:Schneider\} \newline
	\{id:275{\agap},op:S,sn:{\agap}Han{\agap}{\agap}{\agap}{\agap}{\agap}\} \newline
	\{id:490{\agap},op:S,sn:Grundy{\agap}{\agap}{\agap}\} \newline
	}}
	\end{minipage}
}

\newcommand{\alignedArqs}{
	\begin{minipage}[b]{.45\linewidth}
	\texttt{\footnotesize{
	\{id:24{\agap}{\agap},op:A,sn:Schne{\agap}{\agap}{\agap}{\agap}{\agap}ider{\agap}{\agap},mo{\agap}bil{\agap}{\agap}{\agap}e:123456\}
	\{id:2487,op:A,sn:W{\agap}{\agap}{\agap}{\agap}{\agap}{\agap}{\agap}{\agap}{\agap}{\agap}{\agap}{\agap}{\agap}{\agap}{\agap}{\agap}{\agap}{\agap}{\agap}{\agap}il{\agap}{\agap}{\agap}l{\agap}{\agap}{\agap}{\agap}{\agap}{\agap}{\agap}\}
	\{id:3106,op:A,sn:Hi{\agap}ne,gn:Cameron,postalCode:3{\agap}3589\}
	}}
	\end{minipage}
}

\vspace{-0.3cm}
\begin{figure}[ht]%
	\begin{minipage}[b]{0.5\textwidth}{\alignedSrqs}\\
	\vspace{-.7cm}
	\subcaption{\small Cluster 1 alignment} \label{chap7fig:reqalignment1}%
	\end{minipage}%
	\\
	\begin{minipage}[b]{0.5\textwidth}{\alignedArqs}\\
	\vspace{-.7cm}
	\subcaption{\small Cluster 2 alignment} \label{chap7fig:reqalignment2}%
	\end{minipage}%
\vspace{-.2cm}
\caption{MSA results of the requests in tables \ref{tab:searchcluster} and \ref{tab:addcluster}.}%
\label{chap7fig:alignment}%
\end{figure}

\subsubsection{Formulating the Request {\Prototype}}
\label{subsec:consensusseqcal}

Having derived the MSA profile for the request messages of each cluster, the next step is to extract the common features from the MSA profile into a single character sequence, which we call the \emph{request {\prototype}}, to facilitate efficient runtime comparison with an incoming request message. 



\begin{figure*}[ht]
     \centering
     \includegraphics[width=15cm, height=3cm]{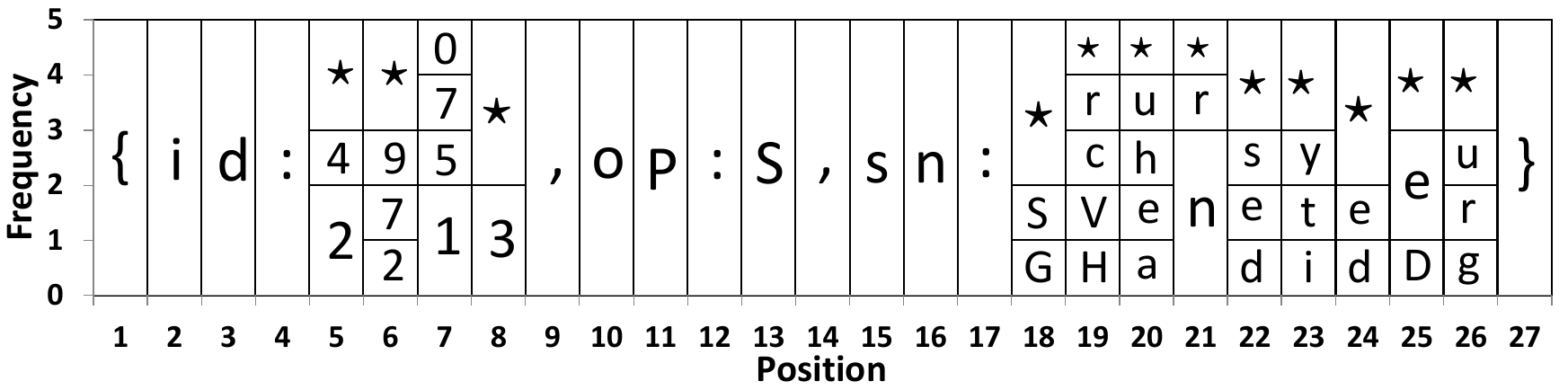}
	 \vspace{-.2cm}
     \caption{Character Frequencies in the Alignment Result of Search Requests.}
     \label{chap7tab:charfreq}
\end{figure*}

From all the aligned request messages in a cluster, we derive a byte (or character) occurrence count table.
Figure~\ref{chap7tab:charfreq} graphically depicts byte frequencies at each position for the example alignment in Figure~\ref{chap7fig:alignment}. Each column represents a position in the alignment result.  The frequencies of the different bytes occurring at each position are displayed as a stacked bar graph.

Based on the byte occurrence table, we formulate the \emph{request {\prototype}} by extending the concept of a \emph{consensus sequence}~\cite{ConsensusComparison} commonly used in summarising a MSA profile.  A consensus sequence can be viewed as a sequence of \emph{consensus symbols}, where the consensus symbol $c_i$ is the most commonly occurring byte at the position $i$. In our extension, a request {\prototype}, $\mathbf{p}$, is calculated by iterating each byte position of the MSA profile, to calculate a \emph{prototype symbol}, $p_i$, at each position, according to Equation~\ref{eq:consensus}

\vspace{-.4cm}
\begin{equation}
\label{eq:consensus}
    p_i = \left\{ \begin{array}{rl}
                 c_i &\mbox{ if $q(c_i) \geq f \land c_i \neq \gap$} \\
                 \perp &\mbox{ if $q(c_i) \geq \frac{1}{2} \land c_i = \gap$} \\
                 {\wildcard} &\mbox{ otherwise }
    \end{array}
        \right.
\end{equation}

where $q(c_i)$ denotes the relative frequency at position $i$ of the consensus
symbol $c_i$, $f$ the relative frequency threshold, `$\gap$' a gap, `$\wildcard$' the `wildcard' symbol, and `$\perp$' represents a truncation. After calculating the prototype symbol for each position, any truncation symbols are then deleted from the request {\prototype}.

Introducing wildcards and truncations into the {\prototype}s allows us to
distinguish between gaps and where there is no consensus. If the relative
frequency $q(c_i)$ is at or above the threshold $f$, we insert the consensus
symbol into our prototype (unless the consensus symbol is a gap). If the
relative frequency is below the threshold, then we insert a wildcard. If the
consensus symbol is a gap and it is in the majority, then we leave that
position as empty ({\em \ie} deleted). Wildcards allow us to encode where
there are high variability sections of the message. Our experiments have shown
that without truncation, consensus sequences became artificially long as there
are generally many gaps. By truncating the gaps, the lengths of the prototypes
become similar to the typical lengths of messages in the cluster. The
{\prototype} for a cluster of request messages can differentiate stable
positions from variant positions. Moreover, it can identify consensus symbols
that can be utilised for matching.

Applying our request {\prototype} method, using a frequency threshold $f=0.8$, to the example clusters in tables \ref{tab:searchcluster} and \ref{tab:addcluster} yield the following results:

    \hspace{-0.1cm} \textbf{Request prototype for the search cluster:}

    \hspace{0.2cm}\texttt{\footnotesize{\{id:${\wildcard}{\wildcard}{\wildcard}$,op:S,sn:${\wildcard}{\wildcard}{\wildcard}{\wildcard}{\wildcard}{\wildcard}{\wildcard}$\}}}

    \hspace{-0.1cm} \textbf{Request prototype for the add cluster:}

    \hspace{0.2cm}\texttt{\footnotesize{\{id:${\wildcard}{\wildcard}{\wildcard}{\wildcard}$,op:A,sn:${\wildcard}{\wildcard}{\wildcard}{\wildcard}{\wildcard}{\wildcard}{\wildcard}{\wildcard}{\wildcard}{\wildcard}{\wildcard}{\wildcard}{\wildcard}{\wildcard}$l${\wildcard}{\wildcard}{\wildcard}{\wildcard}{\wildcard}{\wildcard}{\wildcard}$\}}}

\noindent
Please note that the add prototype contains an `l' from coincidentally aligning `l's from `mobile', `Will' and `postalCode'.

\smallskip

\subsubsection{Deriving Entropy-Based Positional Weightings}
\label{ss:weightings}

The final substep of the offline analysis is to estimate the importance
of different sections of the {\prototype} to be used as weights
during the matching at runtime. Our aim is to give a greater weight
to the bytes that correspond to the operation type.
We make use of the observation that structure information (such as operation type)
is more stable than payload information. We use entropy, as defined by 
the Shannon Index \cite{shannon:1948}, as an estimation of instability.
Using the aligned messages for each cluster, we calculate the inverse
entropy at each byte position aligned to the prototype, to derive
a weightings array, $\mathbf{w}$, of the same length as the prototype $p$.
Full details of the calculation method are given in \cite{Du:2015}.

\subsection{Live Request Runtime Matching (Step R4)}
\label{ss:runtimematching}

At runtime, the formulated request {\prototype}s are used to match
incoming requests from the system under test.
We adapt the Needleman-Wunsch algorithm to calculate the
matching distance between an incoming request and the request {\prototype} for each operation type (or cluster).
We modify the Needleman-Wunsch scoring function, $S$ (see equation~\ref{eq:wildcardnw}), by giving a special score for alignments with wildcard characters and multiplying all scores by their corresponding positional entropy weights (from equation~\ref{eq:wildcardnw}).

\vspace{-.4cm}
\begin{equation}
\label{eq:wildcardnw}
S({p}_i,{r}_j) = \left\{ \begin{array}{ll}
		 w_i\! \cdot\! m &\mbox{ if ${p}_i = {r}_j \land {p}_i \neq \wildcard $} \\
		 w_i\! \cdot\! d &\mbox{ if ${p}_i \neq {r}_j \land {p}_i \neq \wildcard $} \\
		 w_i\! \cdot\! x &\mbox{ if ${p}_i = \wildcard $}
    \end{array}
	\right.
\end{equation}
where ${p}_i$ denotes the $i$th character in the {\prototype},
${r}_j$ the $j$th character in the incoming request,
$w_i$ the weighting value (\cf~\ref{ss:weightings}), and
$m$ and $d$ denote constants of the Needleman-Wunsch identical score and difference penalty, respectively,
and finally $x$ denotes the wildcard matching constant. 
In our experiments we used standard values for the Needleman-Wunsch
constants: $m = 1$, $d = −1$ and used $x = 0$.

Using the modified scoring equation~\ref{eq:wildcardnw}, we apply Needleman-Wunsch to align the consensus prototype with an incoming request, calculating an absolute alignment score, $s$.

The relative distance, denoted as $d_{\mathrm{rel}}$, is calculated from the
absolute alignment score to normalise for consensus prototypes of different
lengths, different entropy weights and different numbers of wildcards.  The
relative distance is in the range 0 to 1, inclusive, where 0 signifies the
best possible match with the consensus, and 1 represents the furthest possible
distance. It is calculated according to Equation~\ref{eq:reldist}:

\vspace{-0.4cm}
\begin{equation}
\label{eq:reldist}
d_{\mathrm{rel}}(\mathbf{p}, \mathbf{r}) =
	1 - \frac{s(\mathbf{p}, \mathbf{r}) - s_{\mathrm{min}}(\mathbf{p}) }
		     {s_{\mathrm{max}}(\mathbf{p} - s_{\mathrm{min}}(\mathbf{p})}
\end{equation}

where $s_{\mathrm{max}}$ denotes the maximum possible alignment score
and $s_{\mathrm{min}}$ the minimum possible alignment score for the given
{\prototype} as defined by equations \ref{eq:maxscore} and \ref{eq:minscore},
respectively.

\vspace{-0.4cm}
\begin{equation}
\label{eq:maxscore}
{s_{\mathrm{max}}(\mathbf{p})} = \sum_{i=1}^{|\mathbf{p}|} S(\mathbf{p}_i,\mathbf{p}_i)
\end{equation}

\vspace{-0.4cm}
\begin{equation}
\label{eq:minscore}
{s_{\mathrm{min}}(\mathbf{p})} = \sum_{i=1}^{|\mathbf{p}|} S(\mathbf{p}_i,\varnothing)
\end{equation}

$\varnothing$ denotes a special symbol different to all characters in
the {\prototype}.

The {\prototype} with the least distance to the incoming message is
selected as the matching prototype, therefore identifying the matching
transaction cluster.

\smallskip

As an example, consider an incoming add request with the byte sequence
\texttt{\footnotesize{\{id:37,op:A,sn:Durand\}}}.
Aligning the request against the search {\prototype} yields

\begin{tabular}{ll}
\emph{request:} & \texttt{\footnotesize{\{id:3\agap7,op:A,sn:{\agap}Durand\}}} \\
\emph{prototype:}        & \texttt{\footnotesize{\{id:???,op:S,sn:???????\}}} \\
\end{tabular}

\smallskip \noindent
Using Equation~\ref{eq:reldist}, the weighted relative distance is calculated
to be 0.0715.
In comparison, the add consensus prototype produces the alignment below with a relative distance of 0.068.

\begin{tabular}{ll}
\emph{request:}   & \hspace{-.4cm} \texttt{\footnotesize{\{id:37{\agap}{\agap},op:A,sn:{\agap}{\agap}{\agap}{\agap}{\agap}{\agap}{\agap}{\agap}D{\agap}u{\agap}r{\agap}an{\agap}d{\agap}{\agap}{\agap}\}}} \\
\emph{prototype:} & \hspace{-.4cm} \texttt{\footnotesize{\{id:????,op:A,sn:?????????????l???????\}}} \\
\end{tabular}

\smallskip \noindent
Consequently, the add prototype is the closest matching, causing the add cluster to be selected.

\subsection{Response Transformation (Step R5)}
\label{sec:transform}

The final step in our approach is to send a customised response for the
incoming request by performing some dynamic substitutions on a response
message from the matched cluster. Here we use the symmetric field technique
described in \cite{Du:2013}, where character sequences which occur in both the
request and response messages of the chosen transaction are substituted with
the corresponding characters from the live request in the generated
response. We take the response from the centroid transaction
\cite{Du:2013SoftMine} of the selected cluster and apply the symmetric field
substitution.

In our example, the centroid transaction from the add cluster is given
below. There is one symmetric field (boxed).

\begin{tabular}{ll}
\textit{request:} & \hspace{-.4cm} \texttt{\footnotesize{\boxed{\texttt{\{id:24,op:A}},sn:Schneider,mobile:123456\}}} \\
\textit{response:} & \hspace{-.4cm} \texttt{\footnotesize{\boxed{\texttt{\{id:24,op:A}}ddRsp,result:Ok\}}} \\
\end{tabular}

\smallskip \noindent After performing the symmetric field substitution the
final generated response is
\texttt{\footnotesize{\{id:37,op:AddRsp,result:Ok\}}}





\section{Experimental Evaluation}
\label{sec:evaluation}

We wanted to assess two key characteristics of our request prototype approach: \emph{accuracy} and \emph{efficiency}, aiming to answer the following research questions:
\begin{enumerate}
\item \textbf{RQ1} (\textbf{Accuracy}): having \emph{request
    {\prototype}s} formulated at the pre-processing stage, is our
  approach able to generate accurate, protocol-conformant responses?
\item \textbf{RQ2} (\textbf{Efficiency}): is our technique efficient enough to
  generate timely responses, even for large transaction libraries?

\end{enumerate}

A more fine-grained evaluation of the approach and its sensitivity to
parameter values used in the algorithms is available in \cite{Du:2015}. 


\subsection{Case Study Protocols and Traces}
\label{ss:casestudies}

\begin{table}[h]
\centering
\resizebox{0.49\textwidth}{!}{
\begin{tabular}{|l|c|c|c|c|}
  \hline
  Protocol & Binary/Text & Fields & \#Ops. & \#Transactions \\
  \hline \hline
  IMS & binary & fixed length & 5 & 800 \\ \hline
  LDAP & binary & length-encoded & 10 & 2177 \\ \hline
  SOAP & text & delimited & 6 & 1000 \\ \hline
  Twitter (REST) & text & delimited & 6 & 1825 \\ \hline
\end{tabular}}
\vspace{-.1cm}
\caption{Message trace datasets. 
}
\label{tab:sampleset}
\end{table}

In order to answer these questions, we applied our technique on message
trace datasets from 4 case study protocols:
\emph{IMS}~\cite{imsredbook} (a binary mainframe protocol),
\emph{LDAP}~\cite{ldap} (a binary directory service protocol),
\emph{SOAP}~\cite{soap} (a textual protocol, with an Enterprise Resource
Planning (ERP) system messaging system services), and
\emph{RESTful Twitter}~\cite{twitter} (a JSON protocol for the Twitter social
media service). We chose these 4 protocols because: (i) they are widely used in enterprise environments, (ii) they
represent a good mix of text-based protocols (SOAP and RESTful Twitter) and
binary protocols (IMS and LDAP), (iii) they use either fixed length, length encoding or
delimiters to structure protocol messages,\footnote{Given a protocol message,
  length fields or delimiters are used to convert its structure into a
  sequence of bytes that can be transmitted over the wire. Specifically, a
  length field is a number of bytes that show the length of another field,
  while a delimiter is a byte (or byte sequence) with a
  known value that indicates the end of a field.} and (iv) each of them includes a diverse
number of operation types, as indicated by the \emph{Ops} column. The number
of request-response interactions for each test case is shown as column
\emph{\#Transactions} in Table~\ref{tab:sampleset}. 

Our message trace datasets are available for download
at {\sf {\small { http://quoll.ict.swin.edu.au/doc/message\_traces.html}}}

\begin{table}[t]
\centering
\resizebox{0.49\textwidth}{!}{
\begin{tabular}{|l|l|}
\hline
Incoming Request & \{id:15,op:S,sn:Du\} \\
\hline
Expected Response & \{id:15,op:SearchRsp,result:Ok,gn:Miao,sn:Du\} \\
\hline
\multirow{2}{*}{Valid Responses} & \{id:15,op:SearchRsp,result:Ok,gn:Miao,sn:Du\} \\\cline{2-2}
& \{id:15,op:SearchRsp,result:Ok,gn:Menka,sn:Du\} \\ 
\hline
\multirow{2}{*}{Invalid Responses} & \{id:15,op:\textbf{\emph{AddRsp}},result:Ok\} \\\cline{2-2}
& \{id:15,op:SearchRsp,result:Ok,gn:Miao\textbf{\emph{\},sn:Du}} \\ 
\hline
\end{tabular}
}
\vspace{-.1cm}
\caption{Examples of valid and invalid emulated responses.}
\label{tab:validexample}
\end{table}


\subsection{Compared Techniques}

We compared the proposed {\Prototype} method with two other methods. The
baseline for comparison was a hash lookup. If the hash code of an incoming
request matched the hash code of a request in the transaction library, then
the associated response was replayed (without any transformation). This
approach can only work when a request identical to the live request occurred
in the recording. It is a standard record-and-replay approach used for
situations where nothing is known about the protocol. Our second compared
technique is the \textit{Whole Library}~\cite{Du:2013} approach, an earlier
version of opaque service virtualisation which does not have an analysis
phase. It compares an incoming request to the raw requests in the entire
transaction library, and for the nearest matching request, the associated
response is transformed, using the same method as described in
Section~\ref{sec:transform}. Its weaknesses are that it cannot generate
responses in real-time for large transaction libraries, and it is susceptible
to sending back responses of the wrong operation type if the transaction
library contains a recorded request of a different operation but with similar
payload information.


\subsection{Accuracy Evaluation (RQ1)}
\label{ssb:effectiveness}

\subsubsection{Methodology}


Cross-validation 
is a popular model validation
method for assessing how accurately a predictive model will perform in
practice. For the purpose of our evaluation, we applied the commonly used
10-fold cross-validation approach \cite{mclachlan:2004} to
all 4 case study datasets.

We randomly partitioned each of the original interaction datasets into 10 groups. Of
these 10 groups, one group
is considered to be the {\em evaluation group} for testing our approach, and
the remaining 9 groups constitute the {\em training set}. This process is then
repeated 10 times (the same as the number of groups), so that each of the 10
groups will be used as the evaluation group once.
When running each experiment with each trace dataset, we applied our approach to each request message in
the {\em evaluation group}, referred to as the {\em incoming request}, to generate an
{\em emulated response}. 
The entire cross validation process was repeated ten times for each
experiment using different random seeds.

Having generated a response for each incoming request, we then
used a validation script to assess the accuracy.
The script used a protocol decoder
to parse the emulated response and then compared it to the original
recorded response (the \emph{expected response}) from the evaluation group. 
The emulated response was classified as \emph{valid} or \emph{invalid}, according to the following
definitions:
\begin{enumerate}
\item \textbf{\em Valid:} the emulated response conformed to the protocol message format
(\ie was successfully parsed by the decoder)
and the operation type of the emulated response was the same as the expected response.
Note that contents of the emulated response payload may differ to the 
expected response and still be considered valid.
\item \textbf{\em Invalid:} the emulated response was not structured according
to the expected message format of the protocol, or the operation type of
the emulated response was different to the expected response.
\end{enumerate}

Table~\ref{tab:validexample} shows some example valid and invalid responses to
a given request.

\smallskip

\subsubsection{Results}

Table~\ref{tab:accuracy} summarises the accuracy of the consensus prototype
approach versus the whole library approach for the 4 datasets evaluated. We
defined accuracy as the ratio of valid responses generated versus the total
size of the dataset.

The {\Prototype} approach has a high accuracy for all 4 protocols tested.  It
produces 100\% accuracy for the IMS and SOAP datasets, and greater than 99\%
accuracy for LDAP and Twitter. The {\Prototype} approach is also significantly
more accurate than the Whole Library approach for two of the four protocols
tested (IMS and LDAP). It is equally accurate for the SOAP dataset (both
100\%) and marginally less accurate for Twitter. Overall, the {\Prototype}
approach outperforms the Whole Library approach with regards to accuracy. This
is despite the latter approach accessing all the available data points in the
recording when matching requests, whereas the {\Prototype} method only makes
use of a compressed representation during the matching process. The reason for
the higher accuracy is that the {\prototype} abstracts away the message
payload information sections (using wildcards), so is less susceptible to
matching a request to the wrong operation type but with the right payload
information, whereas the \emph{Whole Library} approach is susceptible to this
type of error (note the well-formed but invalid responses for the \emph{Whole
  Library} approach in Table~\ref{tab:accuracy}). The hash lookup approach
had low accuracy: although it produced 50\% accuracy for the IMS dataset, this
was only for the IMS acknowledgements as these messages are all identical.

\begin{table}
\centering
\begin{tabular}{|c|c|c|c|}
  \hline
   & Hash Lookup & Whole Library & Prototype \\\hline\hline
  IMS & 50.0\% & 75.25\% & 100\% \\ \hline
  LDAP & 5.36\% & 94.12 \% &  99.95\% \\ \hline
  SOAP & 0.5\% & 100\% &  100\% \\ \hline
  Twitter & 30.1\% & 99.56\% &  99.34\% \\ \hline
\end{tabular}
\vspace{-.1cm}
\caption{Response accuracy of opaque service emulation}
\label{tab:accuracy}
\end{table}

\subsection{\textbf{Efficiency (RQ2)}}
\label{ssb:efficiency}

\subsubsection{Methodology}

We instrumented the response generation calls made
during the cross validation evaluation 
to log the response times. We then calculated
the average response generation time for each approach
and protocol.
Tests were run on an
Intel Xeon E5440 2.83GHz CPUs with 24GB of main memory available.
For comparison we also logged the actual response generation
times of the real services during the message trace recording
process.

\begin{table}[t]
\centering
\scriptsize
    \begin{tabular}{|>{\centering\arraybackslash}m{0.8cm} |c|>{\centering\arraybackslash}m{1.1cm}| >{\centering\arraybackslash}m{1.1cm} |>{\centering\arraybackslash}m{1.1cm} |>{\centering\arraybackslash}m{1cm}|}
        \hline
       &\scriptsize{No.}& \scriptsize{Hash Lookup} & \scriptsize{Whole Library} & \scriptsize{Prototype} & \scriptsize{Real System} \\\hline\hline
      IMS  &800& 0.11 & 470.99 & 3.24& 518\\ \hline
      LDAP &2177& 0.11 & 835.91 & 2.88 & 28 \\ \hline
      SOAP &1000& 0.11 & 380.24 & 3.35 & 65 \\ \hline
      Twitter&1825& 0.11 & 464.09  & 36.62 & 417 \\ \hline
    \end{tabular}
\caption{Average total response generation time (ms)}
\label{tab:efficiency}
\end{table}

\smallskip

\subsubsection{Results}

Table \ref{tab:efficiency} compares the average response generation time of
the Whole Library and {\Prototype} approaches. Also shown are average response
times of the real services. The results show that the {\Prototype} approach is
very efficient at generating responses. The approach is about two orders of
magnitude faster than the Whole Library approach. The {\Prototype} approach is
also significantly faster than the real services being emulated. This is
crucial for supporting testing of an enterprise system under test under
realistic performance conditions (delays can be added to slow down the
emulated response, but not the other way around.)

It should be noted that the LDAP and SOAP services ran on the local network,
whereas IMS and Twitter were remote services. 
To compensate for the remote access network latency, we subtracted this (around 15
milliseconds) from the response times. Even allowing for network
latency, the {\Prototype} method was still faster than the real services. For comparison, we
included response times of the hash lookup method. The response retrieval time
was near instant (less than 0.005 milliseconds). Local network latency was the
major component of total response generation time, about 0.11 milliseconds.

\section{Product Integration}
\label{sec:architecture}

Opaque service virtualisation has recently been integrated into CA
Technologies' commercial product: CA Service Virtualization, as a new feature
named ``Opaque Data Processing'' (ODP).  ODP was initially included in CA
Service Virtualization version 8.0, released in January 2015, and has been
sold to customers \cite{theaustralian}. The implementation of ODP in version
8.0 followed the Whole Library approach \cite{Du:2013}. The {\prototype}
approach proposed in this paper was released in version 9.1 of CA
Service Virtualization. We describe some of the implementation details to
industrialise the approach and key lessons learned.

\subsection{Implementation Architecture}

\begin{figure}
\centering
\includegraphics[width=8.6cm]{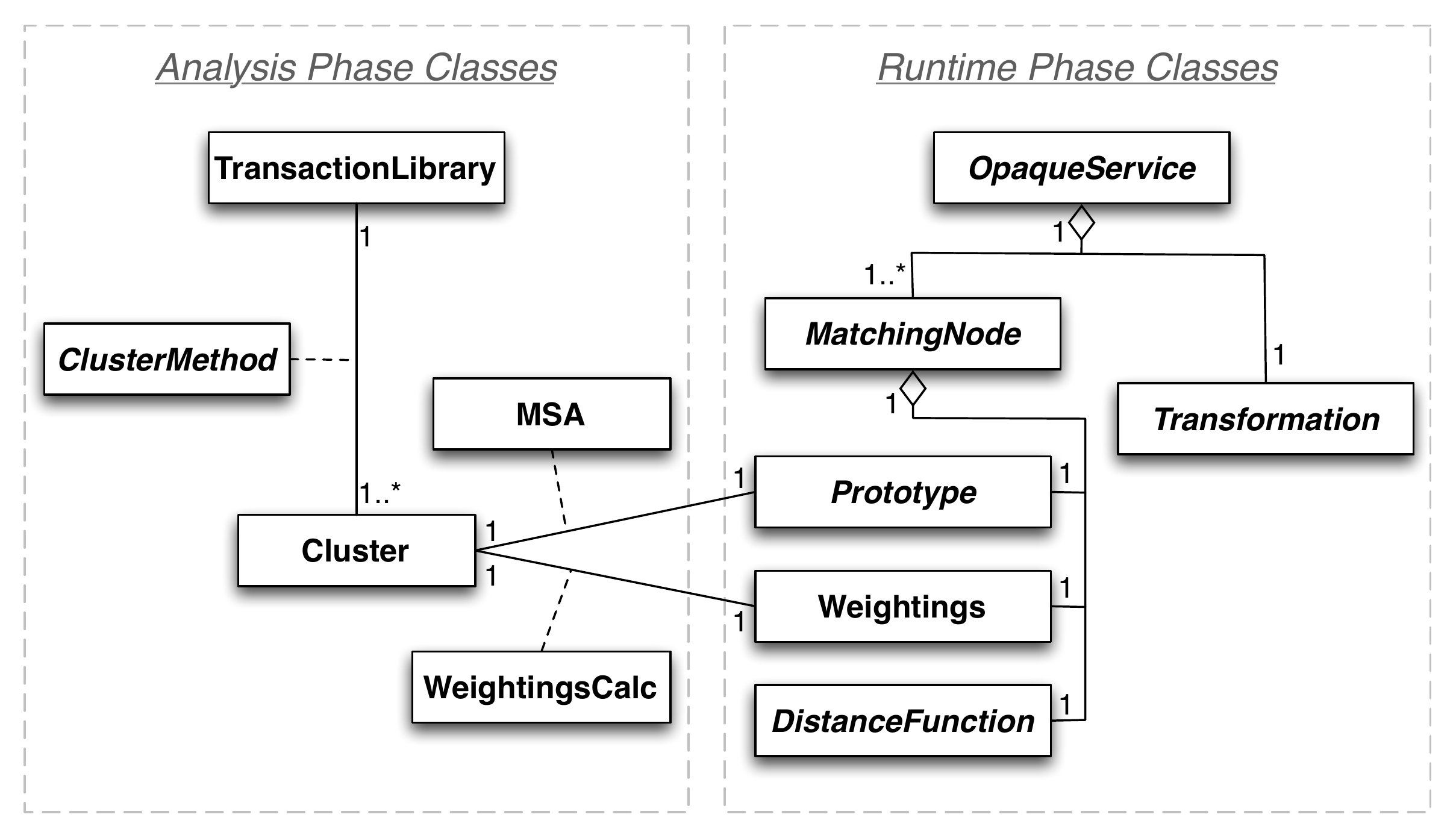}
\caption{UML class diagram of the ODP `Bilby' implementation}
\label{fig:umlbilby}
\end{figure}

We implemented our approach (see section~\ref{sec:approach}) as a Java package
codenamed `Bilby'. Due to commercial agreements we are unable to release the
source code but we will describe the key aspects of the implementation.
Figure~\ref{fig:umlbilby} shows a UML diagram of the key Bilby classes. The
implementation consists of two parts: classes to support the analysis phase
and classes to support the runtime (playback) phase.

For the analysis phase, the transactions collected during the network
recording are stored in an \emph{TransactionLibrary} class. The clustering is
done via a \emph{ClusterMethod}, returning a list of \emph{Cluster}
classes. Each cluster is then used to derive a \emph{Prototype} and a
\emph{Weightings} class, via the algorithms encoded in the \emph{MSA} and
\emph{WeightingsCalc} classes, respectively.

For the runtime phase, the top level interface is given by an \emph{OpaqueService}
class, which is composed of a list of \emph{MatchingNode}s and a
\emph{Transformation} class. A \emph{MatchingNode} is realised by aggregating a
\emph{Prototype} and the associated \emph{Weightings} with a
\emph{DistanceFunction} (the method for calculating the distance between two
messages).

\begin{figure}
\centering
\includegraphics[width=8.6cm]{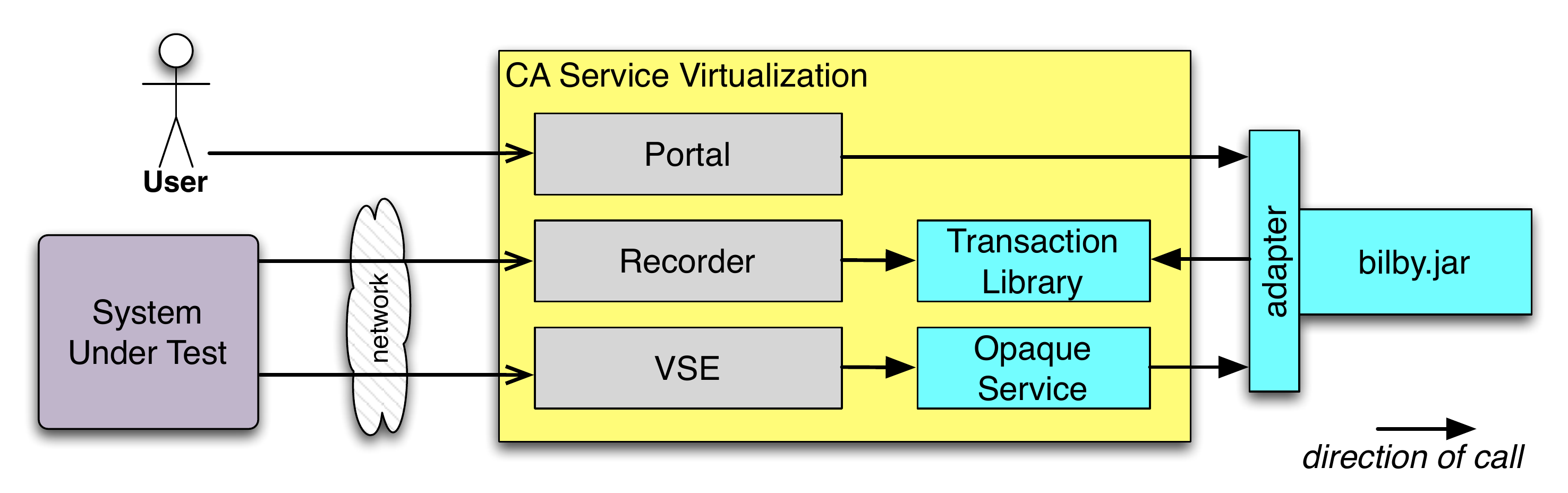}
\caption{Integration of CA Service Virtualization and Bilby}
\label{fig:integration}
\end{figure}

The Bilby package is a jar (Java Archive) and is integrated into the
CA Service Virtualization product. 
An adapter was written to convert between
the internal data structures used by CA Service Virtualization and
Bilby, specifically for requests and responses.
CA Service Virtualization has a component called the Virtual Service Environment (VSE),
which is a platform for deploying virtual service models. The VSE
routes all the network communications between external systems (i.e. systems-under-test)
and the service models. The VSE supports different kinds
of service models. A new type of VSE model was written to support opaque services,
which acts as a wrapper to invoke the Bilby classes.
Figure~\ref{fig:integration} schematically depicts the key elements of
the integration.

\begin{figure}
\centering
\includegraphics[width=8.6cm]{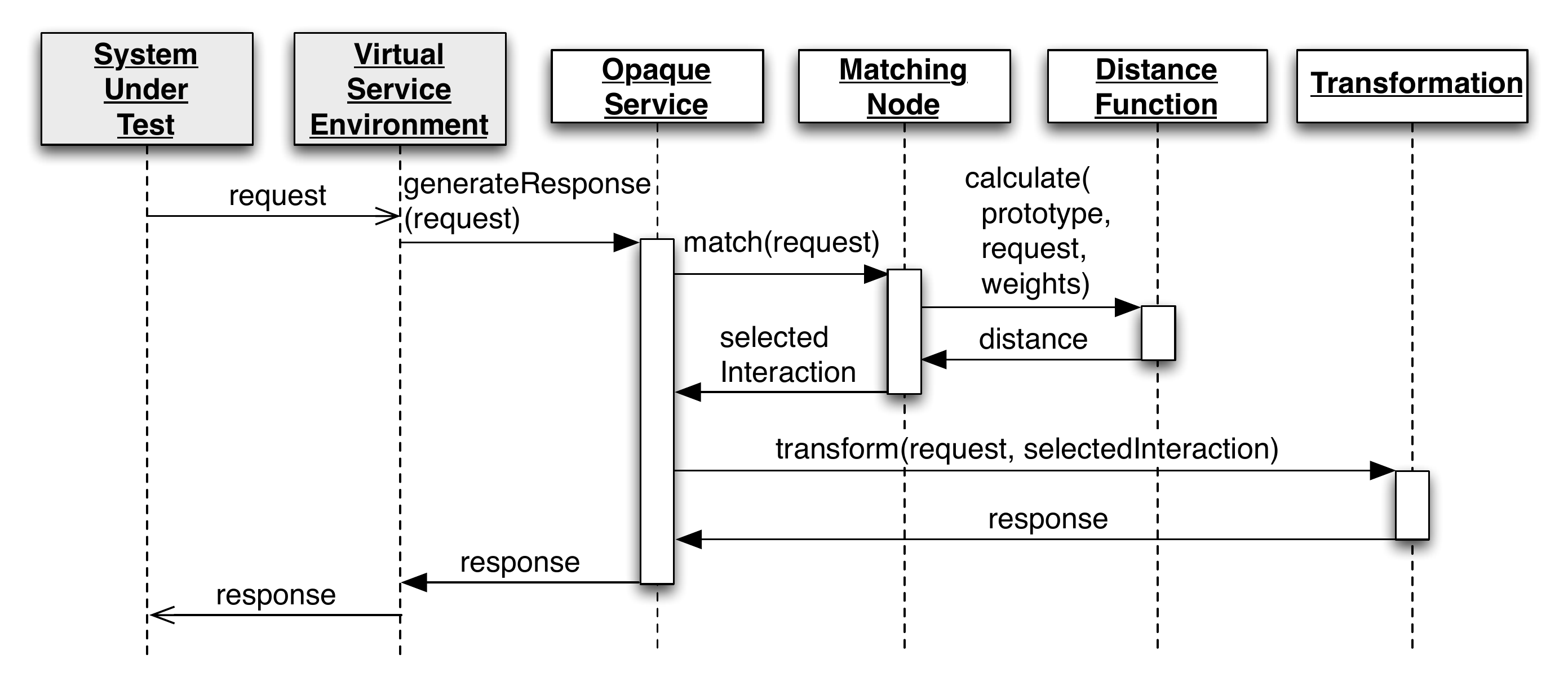}
\caption{UML sequence diagram of ODP playback}
\label{fig:umlsequence}
\end{figure}

Figure~\ref{fig:umlsequence} depicts a UML sequence diagram of a playback event.
A request from the system-under-test is received by the VSE. The VSE invokes
the \emph{OpaqueService} class to generate a response. The \emph{OpaqueService}
class queries all of its \emph{MatchingNode}s to find the best matching
{\prototype} and the associated interaction. The \emph{Transformation}
class is then invoked to perform the symmetric field substitutions to
obtain the final response. The response is then passed back up the call stack
and sent back to the system-under-test.

Extending the CA Service Virtualization user interface portal to support opaque
services was the final integration task. Creating an opaque service
has three stages in the user interface: Record, Configure and Save.
Record is where interactions between the target service and a system-under-test
are captured. 
Configure is where the analysis phase is performed. The user
supervises the clustering process and is allowed to tune the parameters
that will be used for message matching. Once the opaque service is saved
it may be deployed in the VSE like any other virtual service.

\subsection{Challenges and Lessons Learned}

Opaque service virtualization invalidated one of the initial assumptions
of the CA Service Virtualization's architecture -- that all protocol handlers
require a parsing step. The CA Service Virtualization architecture
supports multi-layers of protocol handlers: one layer for a transport protocol
handler and one or more layers for data protocol handlers. The
data protocol handler architecture makes the assumption that the message structure
will be parsed and message fields will be extracted.  However, the opaque service virtualization does not
conform to this pattern, as no fields are explicitly extracted from the messages
and the matching is done on the raw bytes.
For the initial integration performed in the 8.0
release, a workaround to the architectural constraints was
to make ODP an extension of the TCP transport protocol handler.
The limitation of this approach was that ODP was explicitly tied to TCP and
could not be used in conjunction with any other protocol handler.
For the 9.0 release integration, the CA Service Virtualization's overall
architecture was rethought to accommodate ODP, allowing it to be combined
with any transport protocol handler.

The user interface for the opaque service virtualization also posed challenges.
ODP deploys some sophisticated algorithms which are not necessarily easy for
users to understand. We therefore carefully considered which of
the algorithms' parameters should be exposed to the user, and had to provide
robust default settings. An example of this was the clustering algorithm used.
Rather than requiring the user to interactively select clusters (as is required with VAT)
we switched to an automated clustering process which only requires the user
to estimate the number of operation types \cite{wang2009}.

Strong collaboration between the research team, the engineering product team
and product management (located in Australia and the United States)
was key to successful productisation. This
required open mindedness on all sides. Monthly teleconferences between the
researchers and product team helped develop a joint understanding of the
practical problems, identify research topics, and facilitate an exchange of
ideas. There is also a big gap between a research prototype and a commercial
quality software application. The initial implementation was redesigned and
rewritten from scratch, jointly by the researchers and software engineers from
the product team. To enable this collaboration a git repository was set up,
accessible to both the university and CA Technologies staff. A representative
of the research team travelled to the product team's development site for key
stages of the integration work.


\section{Three End User Experiences}
\label{sec:usecases}

We illustrate the ODP's utility with three deployment examples from 
end user organisations. The case studies are based on information
provided by technical services staff working at the end user
organisation sites. Due to confidentiality agreements, the
specifics and names of the organisations cannot be given, but we will
describe them in general terms. For the three case studies
we examine the number of transactions recorded and used for validation,
the accuracy and whether or not the service virtualisation was
ultimately successful. Table~\ref{tab:customerstats} summarises
this information.

\begin{table}
\centering
\resizebox{0.49\textwidth}{!}{
\begin{tabular}{|l|>{\centering\arraybackslash}m{1.4cm}|>{\centering\arraybackslash}m{1.4cm}|c|c|c|}
\hline
Protocol & Recording \#Trans. & Validation \#Trans. & \#Ops. & Accuracy & Success \\
\hline
IMS & 10 & 10 & 5 & 100\% & Yes \\
HL7 & 1000 & 1000 & 5  & 100\% & n/a \\
Tuxedo & 18 & 18 & 9 & 83\% & No \\
\hline
\end{tabular}
}
\caption{ODP accuracy when deployed at end user organisation sites.}
\label{tab:customerstats}
\end{table}

The first use case example was at a bank. The bank was undertaking
a service virtualisation project to facilitate rapid application
deployment. One of the dependent systems that needed to be virtualised
was a mainframe application which used the IMS protocol. Although IMS is supported by CA Service
Virtualization, the target dependency system used an extension to IMS
not supported by the data protocol handler.  Without using ODP this would have meant rebuilding the model for this IMS end point at great time and effort.

For the IMS service, communications with the system-under-test were recorded,
with two sample transactions for each of the five operation types.
An Opaque Service model was then created from the recording.
The system-under-test then connected to the opaque virtual service model and sent
a series of requests to the model, which sent back valid responses
to the system-under-test.
ODP enabled a successful virtualisation of the IMS service that would otherwise have been much more time-consuming and difficult.

The second example use case was at a health service provider. The
provider used a domain specific protocol, called HL7 \cite{hl7}, for applications
to communicate to the patient database. There are different versions
of the HL7 protocol, with the healthcare provider using an older version.
HL7 was not supported by CA Service Virtualization. It was also
considered uneconomic to write a specialised data protocol handler for an
older version of a protocol used by only one customer. ODP was
therefore used to create a virtual service model of the target HL7
services. 

HL7 is a textual format which uses special characters to delimit fields and
records. We recorded a sample of 1000 requests and responses and applied a
cross validation experiment (following the method in \cf~\ref{sec:evaluation})
to measure the accuracy of ODP on the HL7 protocol. The experimental results
showed that ODP produced valid responses with 100\% accuracy for the sample
data. These results show that ODP can be used to support protocols not
handled by the built-in data protocol handlers. At the time of writing the
opaque service model has not yet been deployed at the health care provider's
site.

The third example is at a car rental company, which was
undertaking a service virtualisation. One of the target services
used Oracle Tuxedo, a middleware protocol, which CA Service Virtualization
has no built-in data protocol handler for. The technical
staff therefore tried ODP to virtualise the service.
The service used nine different operation types. For each operation
type two sample transactions were recorded. When the system-under-test
connected to the opaque service model, after an initial exchange of messages,
the system-under-test would hang.
An examination of the message logs revealed that the
conversations between the Tuxedo service and system-under-test were invoked bidirectionally,
\ie the Tuxedo service sometimes sent requests to the system-under-test.
During model playback, the system-under-test would hang while waiting
to receive a request from the model service.
ODP does not currently support message bus style communication.
ODP was therefore not successful for this use case.

Overall the analysis of the use of opaque service virtualisation at
end user organisations shows that it can successfully 
build service models where no data protocol handler exists. It is not
successful in all cases. In particular the Tuxedo use case shows that there
is a need to expand functionality to support message bus-style and other middleware protocols in addition
to client-server protocols.

\section{Discussion and Future Work}
\label{sec:discussion}

We have developed an approach, \emph{opaque service virtualisation}, for automatically generating service responses from message traces which requires no prior knowledge of message structure or decoders or message schemas.
Our approach of using multiple sequence alignment to automatically generate {\prototype}s for the purpose of matching request messages is shown to be accurate, efficient and robust.  Wildcards in the prototypes allow the stable and unstable parts of the request messages for the various operation types to be separated.  Rather than using the prototypes directly for strict matching (such as using it as a regular expression) we instead calculate matching distance through a modified Needleman-Wunsch alignment algorithm.  Since we look for the closest matching prototype, the method is robust even if the prototypes are imperfect.  Moreover, this process can match requests which are slightly different to the prototypes, or are of different length to the prototypes.  This allows the system to handle requests which are outside of the cases directly observed in the trace recordings.

Opaque service virtualisation was integrated with a commercial product and has
been used at customer sites, where protocols not supported by existing tools were successfully emulated. 
Protocol models did not need to be manually developed, making it quicker, and more robust, to deploy opaque service virtualisation than using explicitly modelled protocols.

Further challenges need to be addressed. The number of transactions used in recordings at customer sites are typically much smaller than used in our experiments, and this may affect the response accuracy. We will need to retest our approach with smaller transaction libraries and possibly refine our technique accordingly, or propose a method for customers to easily obtain larger recordings. There are also user experience considerations. There is a need to enable human intervention, particularly to correct the automated process where mistakes occur, since some scenarios require 100\% accuracy.  Our approach uses algorithms not easily understandable by end users. A balance needs to be struck between enabling user control versus not overwhelming the user with the algorithm details. The customer use cases also demonstrated the need to support message bus protocols.  Efficiency can also be further improved, particularly for matching long messages. Using an approximate measure of edit distance is an avenue for exploration~\cite{Schneider:15a}.

Our approach does not consider the service state history in formulating responses. Somewhat surprisingly, in practice a stateless model is sufficient in many cases for emulating a deployment environment. For example: (i) when the emulation target service is stateless, or (ii) when the testing scenario does not contain two equivalent requests, requiring different state affected responses, or (iii) where the testing scenario does not require highly accurate responses (\eg performance and reliability testing rather than conformance testing). To address this limitation, an avenue of future exploration is to process mine the operation sequences to discover stateful models and produce stateful responses. Future work will address stateful modelling, such that writes and updates affect future responses. A method for doing this is to use finite state machine models for the service state as well the state of individual data records.

\section{Summary}

We have described a novel technique for emulating large scale deployments of
complex service-based systems. Our \emph{opaque service virtualisation}
technique has been integrated into a commercial product, CA Service
Virtualization, and successfully deployed on customer sites demonstrating its
industrial utility. The growing practice of DevOps needs more tools to
support production-like test conditions at all stages of the software
development lifecycle. Opaque service virtualisation is a promising new tool
to help support emulation of complex service-based environments via more
accurate request/response trace capture and replay with no knowledge of
protocols and no need to construct models of protocols.




\section{Acknowledgments}
This research was supported by ARC grant LP150100892.




\bibliographystyle{IEEEtran}
\bibliography{./Versteeg_seip_2016}

\end{document}